\documentclass[english,a4paper,11pt]{article}
\usepackage[margin=3cm]{geometry}

\usepackage{booktabs}
\usepackage{amsmath}
\usepackage{algorithm}
\usepackage[final]{graphicx}
\DeclareGraphicsExtensions{.jpg,.jpeg,.pdf,.png,.mps}
\usepackage{epsfig}
\usepackage[round]{natbib}
\usepackage{rotating}
\usepackage{color}
\usepackage{xcolor}
\usepackage{colortbl}
\usepackage{hyperref}
\usepackage{csquotes}
\usepackage[toc,page]{appendix}
 \usepackage{threeparttable}
\usepackage[bitstream-charter]{mathdesign}
\usepackage[T1]{fontenc}
\usepackage{lmodern}
\usepackage{footnotebackref}
\title{Evaluating the resilience of ESG investments in European Markets during turmoil periods\\
\begin{large} 
\texttt{Preprint: this version has not been peer-reviewed.}
\end{large} }

\author{$\mathrm{Barbara \ Iannone}^\mathrm{1,*}, \ \mathrm{Pierdomenico \ Duttilo}^\mathrm{2},  
	\  \mathrm{Stefano \ Antonio\ Gattone}^\mathrm{1}$\\ 
	$^\mathrm{1}$\small{\emph{DiSEGS, University "G. d'Annunzio" of Chieti-Pescara, Pescara, Italy}}\\ 
	$^\mathrm{2}$\small{\emph{Department of Statistical Sciences,  University of Padova, Padova, Italy}}\\
    $^\mathrm{*}$\small{\emph{Corresponding author: 	barbara.iannone@unich.it}}
}

\date{}

\begin{document}
	\maketitle
\begin{abstract}

\noindent
This study investigates the resilience of Environmental, Social, and Governance (ESG) investments during periods of financial instability, comparing them with traditional equity indices across major European markets-Germany, France, and Italy. Using daily returns from October 2021 to February 2024, the analysis explores the effects of key global disruptions such as the Covid-19 pandemic and the Russia-Ukraine conflict on market performance. A mixture of two generalised normal distributions (MGND) and EGARCH-in-mean models are used to identify periods of market turmoil and assess volatility dynamics. The findings indicate that during crises, ESG investments present higher volatility in Germany and Italy than in France. Despite some regional variations, ESG portfolios demonstrate greater resilience compared to traditional ones, offering potential risk mitigation during market shocks. These results underscore the importance of integrating ESG factors into long-term investment strategies, particularly in the face of unpredictable financial turmoil.
\hspace{1cm}
\end{abstract}

\section{Introduction}\label{Intro}
Scholars of different disciplines have always investigated the issue of CSR (Corporate Social Responsibility), ethics, and economics. In the past, important business administration scholars recognised the importance of the role played by the changing natural, social, and economic environment in which a company operates. Gino Zappa, the “father” of the Italian School of Business Administration, affirmed: ``\textit{The action of natural and social forces is inversely related to the action of economic forces in a continuous and varied composition}'' \citep{zappa1937reddito}.

In recent years, ESG (Environmental, Social, and Governance) indices have emerged as critical tools for investors looking to integrate sustainability criteria into their financial strategies. The growing importance of ESG indices reflects a paradigm shift in capital allocation, aimed at optimising returns and promoting responsible corporate practices. The effectiveness and resilience of these indices are often tested during periods of market turbulence.

During the past five years, ESG investments in European markets have exhibited considerable volatility, reflecting both broader market dynamics and the evolving landscape of regulations and investor preferences. This volatility has been particularly shaped by global economic uncertainties, such as the COVID-19 pandemic and geopolitical tensions, which have exacerbated price fluctuations. Even before the outbreak of the COVID-19 pandemic, various social, ecological and economic problems have arisen due to the progressive industrial development brought about by globalization. However, ESG investments have frequently demonstrated greater resilience compared to traditional assets, driven by increasing demand for sustainable solutions and regulatory impetus towards more responsible finance. However, the variability in performance, partly attributable to disparities in the adoption of ESG practices across different sectors and countries, underscores the need for rigorous evaluation and customised risk management strategies for such investments. 

The academic literature in business and finance mostly considered the deep changes, continually creating tools to reconcile economic and ethical values. The regulatory changes were also one of the key elements in explaining the decisions and behaviours of financial actors to promote sustainability \citep{Ahlstrom2022,Abate2023}. In the last decades, several tools have risen and, based on these, investors have attempted to link personal and economic values, reaching a great complexity, still present in defining ethical investors and traditional investors. The ``traditional'' investors choose their investments by adopting economic-financial criteria, like the investment duration, the level of return, the degree of risks, liquidity, tax conditions, and so on. In contrast, the ethical investor combines economic performance with measurable social impacts. Among the latest born of finance, as a subset of responsible investing, the Impact Investing, of the Anglo-Saxon matrix, widespread in finance, refers to the strong desire to combine economic performance with measurable social impacts \citep{weber2016,salzmann2013,rizzi2018,michelucci2016}. These types of investments have two different purposes: on the one hand, they finance the pursuit of social benefit, and on the other hand, they achieve financial profit \citep{tekula2016}. Therefore, ethical investment includes other variables, based on ethical, social, and/or environmental principles. The ethical investor cares not only about the level of expected financial return and the risks associated with it, but also about the origin, the nature of the goods or services offered by the company, its location, or the way managers conduct their business according to their procedures \citep{Cowton1999}. Debates about socially responsible investing have been known in the literature for years: while several scholars claim that responsible investing means giving up a portion of profits for ethical reasons \citep{Bauer2007,Bauer2015,FICH201521,Mohamed2017}, other scholars state that behind such choices there is much more; for example, the 17th Sustainable Development Goals in the 2030 Agenda for Sustainable Development confirm that environmental, social, and economic challenges are widely complex \citep{krosinsky2008,Hemerijck2018,Oehmke2021,Friede2019}.

ESG criteria have become an essential component in investment strategies aimed at addressing global sustainability challenges. Previous research has analysed the returns and volatility of ESG indices compared to their traditional counterparts during periods of financial turmoil \citep{mythesis}. In particular, \cite{Duttilo2023} focused on understanding how ESG investments behave in times of market instability, providing insights into their potential role as a stabilising force compared to conventional assets. This analysis spanned multiple markets, including global, US, Europe, and emerging markets. A two-component mixture of generalised normal distribution was used to objectively identify turmoil periods using a Na\"{i}ve Bayes' classifier, and an EGARCH-in-mean model with exogenous dummy variables was applied to capture the impact of these periods. The results indicated that both returns and volatility were significantly influenced by turmoil, with return-risk performance varying between markets. In particular, the European ESG index exhibited lower volatility than its traditional benchmark, while volatility levels were comparable in other markets. Furthermore, the ESG and non-ESG indices differed in terms of turmoil impact, risk premium, and leverage effect.

Building on this previous research, the current study focuses exclusively on the top three European economies—Germany, France, and Italy—over the period from October 2021 to February 2024. By examining these markets during recent global instability, including the Russia-Ukraine conflict, the study aims to assess whether ESG investments in Europe demonstrate enhanced resilience compared to traditional indices, and whether they can serve as a viable risk mitigation strategy during times of crisis.

The rest of the paper is organised as follows. Section \ref{sec.literature} includes the literature review. Section \ref{sec.method} describes the methodology and the data used. Section \ref{sec.results} illustrates the results of the analysis and Section \ref{sec.conclusions} provides a discussion of the results.

\section{Literature review}\label{sec.literature}
In this section, the main literature referring to the investigated topics will be discussed. There has been a paradigm shift characterised by a progressive departure from the exclusive framework of rational deliberation toward the introduction of behavioural elements \citep{Benlemlih2018,Lapanan2018,BROOKS20181,Chatzitheodorou2019}. The ESG investment or  \textit{``... responsible investment is an approach to investment that explicitly acknowledges the relevance to the investor of environmental, social, and governance factors, and of the long-term health and stability of the market as a whole. It recognises that the generation of long-term sustainable returns is dependent on stable, well-functioning, and well-governed social, environmental, and economic systems''} \citep{UniCam2022}. In other terms, it can be defined as an investment that creates long-term social, environmental, and economic value. Therefore, the concept of value must be revisited: the total value created by a company is the result of the value created by customers, suppliers, lenders, employees, and the community. Over time, the concept of corporate value has evolved significantly in both meaning and composition. Initially, it was mainly focused on the maximisation of profit; subsequently, scholars established that it also affected the maximisation of all business capital. They recognised the value of management aimed at increasing company value, as originally stated in Italy by Gino Zappa \citep{zappa1937reddito,zappa1958,zappa1959,zappa1960}, and by his scholars. The company develops and improves itself thanks to the ability to manage and generate new value for itself and its shareholders. But this was not enough; it must be necessary to consider other needs of the context external to the company itself. The horizon of the companies opened, and they look not only internally, but must necessarily deal with the outside world.

In this context, and particularly in the last decades, one of the most important elements to increase the Corporate Value is an intangible asset: the Corporate Reputation. In fact, during this recent complex period, the corporate reputation itself has played a crucial role. According to \cite{Fombrun2013}, \textit{``it is a perceptual representation of past actions and future prospects of a company that describe the overall appeal of the firm to all its key constituents''}. Several studies analyzed the role of firm reputation in different aspects, such as corporate social responsibility (CSR), CSR reporting, investment strategies, international acquisitions, etc. \citep{HAN2022,Miras2020,Fombrun2012,Fombrun2013}. In summary, a good corporate reputation improves the ability to attract new stakeholders and new shareholders in the financial market, and has also become an essential asset in investors' decision-making processes, particularly during times of crisis \citep{schurmann2006reputation,blajer2014corporate,cole2014applying,nawrocki2022importance}. Several researchers focus their studies on the role of firm reputation in different results: for instance, on the engagement of CSR \citep{Rehman2022}, the impact of the reputation of brokerage house (BH) on the performance of investment strategies, on the influence of country reputation differentials on the market reaction to international acquisitions \citep{HAN2022}, on the influence of CSR reporting on corporate reputation \citep{Miras2020}.

In this way, sustainability in business involves the integration of social, environmental, and governance aspects into business strategies and activities, so it becomes imperative that investors and companies systematically incorporate CSR activities into their investment programmes. It represents a key to achieving the goal of realising change, repairing the mistakes that occurred in the years of industrialisation, and radically transforming the negative consequences, such as the phenomenon of Covid-19 and the crisis it generated. Although the specific risk is the result of uncertainty about dividends and the final value of stocks, the market risk is attributable to a combination of economic and socio-political factors or aggregate events that cause periods of turmoil and profound changes on a global scale \citep{Zigrand2014}.

From the above literature review, although there is a large body of studies that finds
that these indices offer investors solid risk management by obtaining positive returns and
promoting economic sustainability, there is no unanimous consensus on the role of the predictive information of ESG indices instead of traditional indices, to prevent turbulence and reduce volatility in the financial markets. We rely on the literature and our first hypothesis is thus as follows: 

\textbf{H1. The performance of ESG indices and that of traditional indices register the same volatility in times of crisis}

There is a new paradigm for reading a crisis, such as the Covid-19 pandemic, as an opportunity to rebalance portfolios and accelerate the shift to a more sustainable world. The investment community recognises the need for managers to respond to urgent pressures but is increasingly focused on the long-term social and environmental impact (Polman, 2020).

Several studies indicate that the stock indices of leading ESG companies in North America and Europe represent a secure investment haven during significant periods of upheaval and crisis. These indices offer investors a means to manage risk and achieve positive returns while promoting economic sustainability \citep{Katsampoxakis2024,Lin2024,risks11100182}. To obtain minor risks in investments, it may be prudent to construct a portfolio composed of ESG-compliant companies, a prudent choice for investors, as it offers relatively superior risk-adjusted returns compared to corresponding market indices. Simultaneously, it reveals another need: to highlight the urgent focus on ESG framework and scores to protect investors from short- and long-term volatilities and economic vulnerabilities \citep{Zanatto2023,Sharma2024}. It is necessary to develop the measurement of ESG within companies that implemented these activities, as this commitment has impacted the results: These dimensions show the most effects to improve resilience during periods of crisis \citep{Garrido2024}. In the medium term, the ESG index can reduce unexpected volatility: it reveals a forecasting power in terms of information for investors and fund managers \citep{capelli2021forecasting}. Designing a portfolio based on companies that comply with ESG could be a prudent choice for investors, as it yields relatively better risk-adjusted returns than the respective market indices.

Starting with the aforementioned literature review, it is possible to assess that most of the literature sustains a positive effect, in terms of managing the risks, provided by the trend of ESG indices. Following the previous studies, we propose the second hypothesis:

\textbf{H2: The performance of ESG indices shows less volatility than traditional indices.}

Financial turmoil periods are difficult or impossible to predict and cause a significant impact on business activities and therefore deep shocks to the stock market as well. These events can bring companies to a complete standstill as they operate in a complex and interconnected context. In the past, the dot-com bubble burst (2000-2002), the global financial crisis (2007-2008), the European sovereign debt crisis (2010-2012), and the last Covid-19 pandemic period are examples of turmoil periods. In the latter case, equity returns responded dynamically to unanticipated changes in pandemic scenarios. Specifically, the effects of Covid-19 on some of the world’s most important stock exchanges, as well as the empirical relationship between Covid-19 waves and volatility of the stock market \citep{Duttilo2021}, revealed a direct negative effect due to the uncertainty effect on global markets and the unpredictability of future scenarios.
In early 2021, critics established that focussing attention on ESG scores is an outdated activity: The time to rely on such ``blunt aggregatio'' had passed. ESG scores are not an ``equity vaccine'' against declining share prices during times of crisis. Explaining stock returns during the Covid-19 period requires considering a combination of accounting-based measures, such as firm liquidity and leverage, financial performance, supply chain management, and internally developed intangible assets, along with company affiliation with industry and traditional market-based measures of equity risk \citep{Demers2021}.

Lastly, in finance, enjoying a good reputation is considered a competitive advantage in terms of attracting shareholders on the financial market: reputation has a positive impact on financial performance, and this, in turn, on reputation. Finally, in the Covid-19 period, having a good corporate reputation might matter, to spread trust. This allows companies to conduct a deeper examination of their activities in pursuit of a better strategic position through social performance \citep{price2017doing}. The concept that ESG investments improve shareholder value, especially in times of crisis, is a topic of considerable debate. Advocates of ESG claim that these investments help build social capital and trust in every company. Their thesis is that responsible behavior, both as social and environmental actions, leads to the building of important links between the company and its stakeholders \citep{demers2021esg}. This goodwill will be particularly useful in periods of crisis because it can help to manage risks, as an insurance-like protection against downside risk \citep{godfrey2009relationship}.

Companies that adopt sustainable practices are more resilient during crises and gain competitive advantages in terms of economic performance and reputation \citep{ahmad2024environmental,chen2023environmental,negara2024impact}. 

According to this last part of the literature review, it is possible to assess that most of the literature supports a dynamic effect in terms of stock returns as a reaction to turbulence. We thus test the following hypothesis:

\textbf{H3: The reaction of stock returns is dynamic, as a consequence of financial turmoil, but shows more dynamicity for ESG indices than for traditional indices.}

Ultimately, various studies, although not definitively, increasingly indicate that companies with robust ESG practices can offer a more secure investment during periods of market uncertainty \citep{fung2024impact}. The ESG index has indeed shown resilience, particularly compared to other indices, helping investors and policy makers in their decision-making processes in the midst of market uncertainties \citep{ghani2024forecasting}. 

A recent study has revealed that the incorporation of ESG factors in risk analysis can make investments more stable and predictable \citep{capelli2021forecasting}. This is not merely a matter of CSR, but a practical approach to improving investment decisions and reducing unforeseen risks. In short, paying attention to ESG factors is essential for those seeking to avoid unexpected losses and improve long-term investment performance.

\section{Data and methodology}\label{sec.method}
The Euro Stoxx 50 index is chosen to identify periods of stability and turmoil, as it effectively represents the European market through the performance of large and midcap stocks. For Germany, France and Italy, data on daily closing prices of traditional and ESG indices have been collected, as presented in Table \ref{tab: rt source}. The study covers the period from October 18, 2021 to February 19, 2024. Due to the limited availability of data for the selected ESG indices and in order to emphasize the impact of the Russia-Ukraine conflict, the chosen time frame deliberately excludes the COVID-19 shock on financial markets in 2020, as well as other earlier crisis periods.

\begin{table}[ht!]
\caption{Selected traditional and ESG indices}
\label{tab: rt source}
\centering 
\resizebox{10cm}{!}{
\begin{tabular}{llll}
\toprule
Index & Ticker & Type & Area\\
\midrule
Euro Stoxx 50 index & STOXX50E & Traditional & Europe \\
DAX 30 & DAX & Traditional & Germany \\
DAX 30 ESG & DAX3ESGK & ESG & Germany\\
CAC 40 & CAC40 & Traditional & France \\
CAC 40 ESG & CAC40ESG & ESG & France \\
FTSEMIB & FTSEMIB & Traditional & Italy\\
MIB ESG & MIBESG & ESG & Italy \\
\bottomrule 
\end{tabular}
}
  \begin{tablenotes}
\centering
\item[]{\footnotesize 
Source: \textit{finance.yahoo.com}, \textit{investing.com}, reference date February 27, 2024.}
\end{tablenotes}
\end{table}

The daily returns of all equity indices under study were calculated using the natural log-difference approach \citep{Wen2020,Duttilo2021,Duttilo2023} $r_{t}=\ln(P_t-P_{t-1})100$, where: $r_{t}$ is the daily percentage return on the equity index at time $t$, $P_{t}$ is the daily closing price of the equity index at time $t$, and $P_{t-1}$ is the daily closing price of the equity index at time $t-1$.

Mixtures of distributions are commonly used to model daily returns and capture key features such as excess kurtosis and skewness. Although normal mixtures are widely applied, they impose restrictive assumptions on the return distribution. The mixture of generalised normal distributions (MGND) overcomes these limitations with a flexible shape parameter $\nu$, which adjusts tail behaviour and peakedness. The probability density function for the MGND model with $K$ components is defined as:
\begin{equation}
f(r_t\lvert\theta) = \sum_{k=1}^K\pi_k \frac{\nu_k}{2\delta_k\Gamma(1/\nu_k)} \exp\left\{ - \left|\frac{r_t - \mu_k}{\delta_k}\right|^{\nu_k} \right\},    
\end{equation}
where $\theta = (\pi_k, \mu_k, \delta_k, \nu_k)$ are the parameters of the mixture: $\pi_k$ represents the mixing weight of the $k$-th component, $\mu_k$ is the location parameter, $\delta_k$ is the scale, and $\nu_k$ controls the shape and tail heaviness. The variance of each component is $\sigma^2_k = \delta_k^2 \Gamma(3/\nu_k) / \Gamma(1/\nu_k)$. To classify returns \citep{Duttilo2023,mythesis} into stable and turmoil periods, the Na\"{i}ve Bayes' classification rule is used, assigning each return to the mixture component with the highest posterior probability, $\max_{k} \pi_k p(r_t\lvert\theta_k)$. Typically, a smaller shape parameter $\nu_k$ or a higher scale parameter $\sigma_k$ indicates a thicker tail (higher kurtosis and volatility), characterising the turmoil periods, while larger $\nu_k$ and smaller $\sigma_k$ correspond to stable periods (mild kurtosis and volatility). 

The EGARCH-in-Mean (EGARCH-M) model complements this approach by capturing volatility dynamics and the leverage effect. The model includes an exogenous dummy variable, $\text{TURMOIL}_t$, which is equal to 1 during turmoil periods and 0 otherwise. The conditional mean equation is given by 
\begin{equation}
r_t = \mu + m_1 \cdot \text{TURMOIL}_t + \phi_1 r_{t-1} + \lambda h_t + \epsilon_t,    
\end{equation} 
where $\mu$ represents the intercept, $m_1$ measures the impact of turmoil on returns, $\phi_1$ captures the autocorrelation, and $\lambda$ reflects the risk premium, showing the relationship between returns and their conditional volatility $h_t$. The conditional volatility equation is given by,
\begin{equation}
\ln(h^2_t) = \omega + v_1 \cdot \text{TURMOIL}_t + \alpha_1 z_{t-1} + \gamma_1(|z_{t-1}| - E[|z_{t-1}|]) + \beta_1 \ln(h^2_{t-1}),    
\end{equation}
where $\omega$ represents the intercept, $v_1$ measures the effect of turmoil on volatility, $\alpha_1$ captures the sign effect (asymmetry), $\gamma_1$ captures the magnitude of shocks, and $\beta_1$ reflects volatility persistence. The residuals, $z_t = \epsilon_t / \sqrt{h^2_t}$, are modelled using the skewed generalised normal distribution to account for the non-normal features of returns. Together, these models offer a comprehensive framework for analysing shifts between stable and turmoil periods in financial markets.

\section{Results}\label{sec.results}
Table \ref{tab:BasicStatistics} in Appendix \ref{app:tables} summarises the basic statistics of the traditional and ESG indices of Germany, France and Italy. In all three countries, traditional indices consistently report higher mean returns than ESG indices, with Italy showing the largest gap (0.0313 for traditional vs. 0.0289 for ESG). Although volatility, measured by standard deviation, is slightly higher for ESG indices in each country, the differences are minimal. Skewness values are similar for both index types, although the ESG indices exhibit slightly more negative skewness in Italy and France, indicating a slight left-leaning distribution, while in Germany, the ESG skewness is slightly positive. Lastly, kurtosis is marginally higher for ESG indices across all countries, suggesting a tendency for more extreme values in the return distribution of ESG portfolios.

Table \ref{tab:StatisticalTestsResults} in Appendix \ref{app:tables} presents the results of various preliminary statistical hypothesis tests. The Jarque-Bera (JB) test indicates that the daily returns do not follow a normal distribution. The Augmented Dickey-Fuller (ADF) test reveals that the null hypothesis of a unit root is rejected. Furthermore, the ARCH-LM test detects the presence of ARCH effects and heteroskedasticity, as the null hypothesis of no ARCH effect is also rejected at the 1\% significance level.

The stable and turmoil components could be identified on the basis of the scale and shape parameters. Looking at the estimated coefficients of the two-component MGND model in Table \ref{tab:MixtureofGND} in Appendix \ref{app:tables}, a few interesting considerations arise. Firstly, the stable component ($\pi_1=0.8502$) is predominant compared to the turmoil component ($\pi_2=0.1498$). Secondly, the estimated MGND model is bimodally asymmetric $\mu_1=0.1141>\mu_2=-0.4408$. Thirdly, the tails of both components intermediate between the Laplace and Normal distributions. On the other hand, the tails of the turmoil component are more extreme than those of the Laplace distribution because $\sigma_2>\sigma_1$.

\begin{figure}[ht]
\centering
\includegraphics[width=1\textwidth]{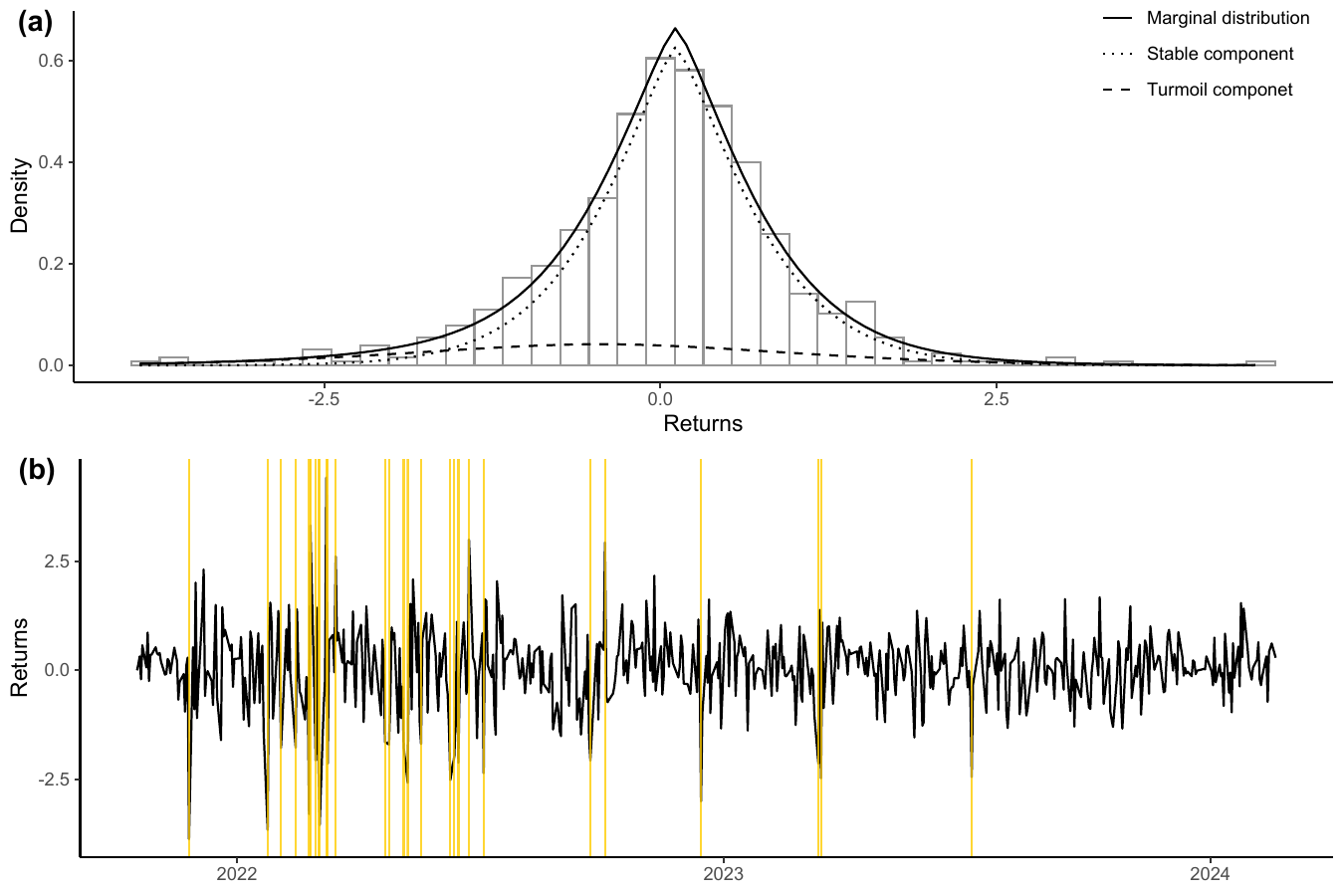}
\caption{Estimated density (panel a) and detected turmoil periods (panel b).}\label{fig.den_turmoil}	
\end{figure}

Figure \ref{fig.den_turmoil}, panel (a), demonstrates that the two-component MGND model effectively captures the heavy tails and leptokurtic nature of the daily returns of STOXX50E. The model also highlights the presence of negative skewness, as indicated by the longer left tail, which corresponds to the turmoil component. Panel (b) illustrates the daily STOXX50E returns, with the yellow vertical lines marking the turmoil periods identified by the two-component MGND model and the Na\"{i}ve Bayes’ classifier. In particular, the Russia-Ukraine conflict stands out as the most volatile period in 2022, while 2023 and 2024 are characterised by relative stability.

\begin{figure}[ht]
\centering
\includegraphics[width=0.8\textwidth]{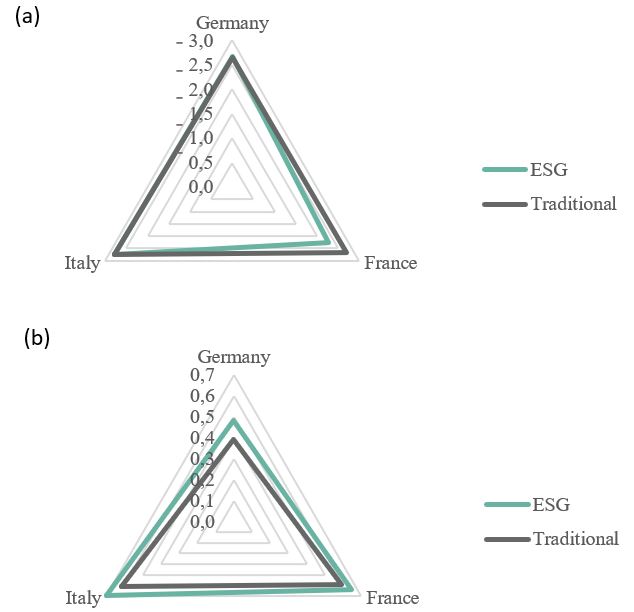}
\caption{Impact of turmoil periods on the conditional mean (panel a) and volatility (panel b) by index type and market.}\label{fig:radarplot}	
\end{figure}

Table \ref{tab:Conditional Mean} in Appendix \ref{app:tables} presents the estimated coefficients for the conditional mean equation. The results indicate that periods of turmoil led to a reduction in the conditional mean, as evidenced by the negative and statistically significant coefficient $m_1$. Figure \ref{fig:radarplot} panel (a) offers a clear visualisation of the effect of these turmoil periods on the conditional mean, broken down by index type and market. The impact is most pronounced in Italy, followed by Germany and France. In particular, no significant differences were observed between the ESG and non-ESG indices, except in France, where the traditional index experienced a greater impact on the conditional mean than its ESG counterpart. Furthermore, the analysis reveals a statistically significant risk premium in all indices. Traditional indices generally exhibit a higher risk premium compared to ESG indices, with the exception of Italy, where the risk premiums are almost identical.

Table \ref{tab:Conditional Variance} in Appendix \ref{app:tables} shows the estimated coefficients for the conditional volatility equation. The results show that the turmoil periods led to an increase in conditional volatility, as indicated by the positive and statistically significant coefficient $v_1$. Figure \ref{fig:radarplot} shows how the magnitude of this impact differs by index type and country market, with the strongest effects observed in Italy, followed by France and Germany. Traditional indices in both Italy and France experienced a more pronounced increase in conditional volatility compared to their ESG counterparts. The statistically significant coefficients $\alpha_1$ and $\beta_1$ indicate the presence of ARCH and GARCH effects. Furthermore, the analysis reveals a negative leverage effect, as reflected by $\alpha_1$ and $\gamma_1$. In general, the effect of leverage is similar between the ESG indices and their respective market benchmarks.

\begin{figure}[H]
\centering
\includegraphics[width=1\textwidth]{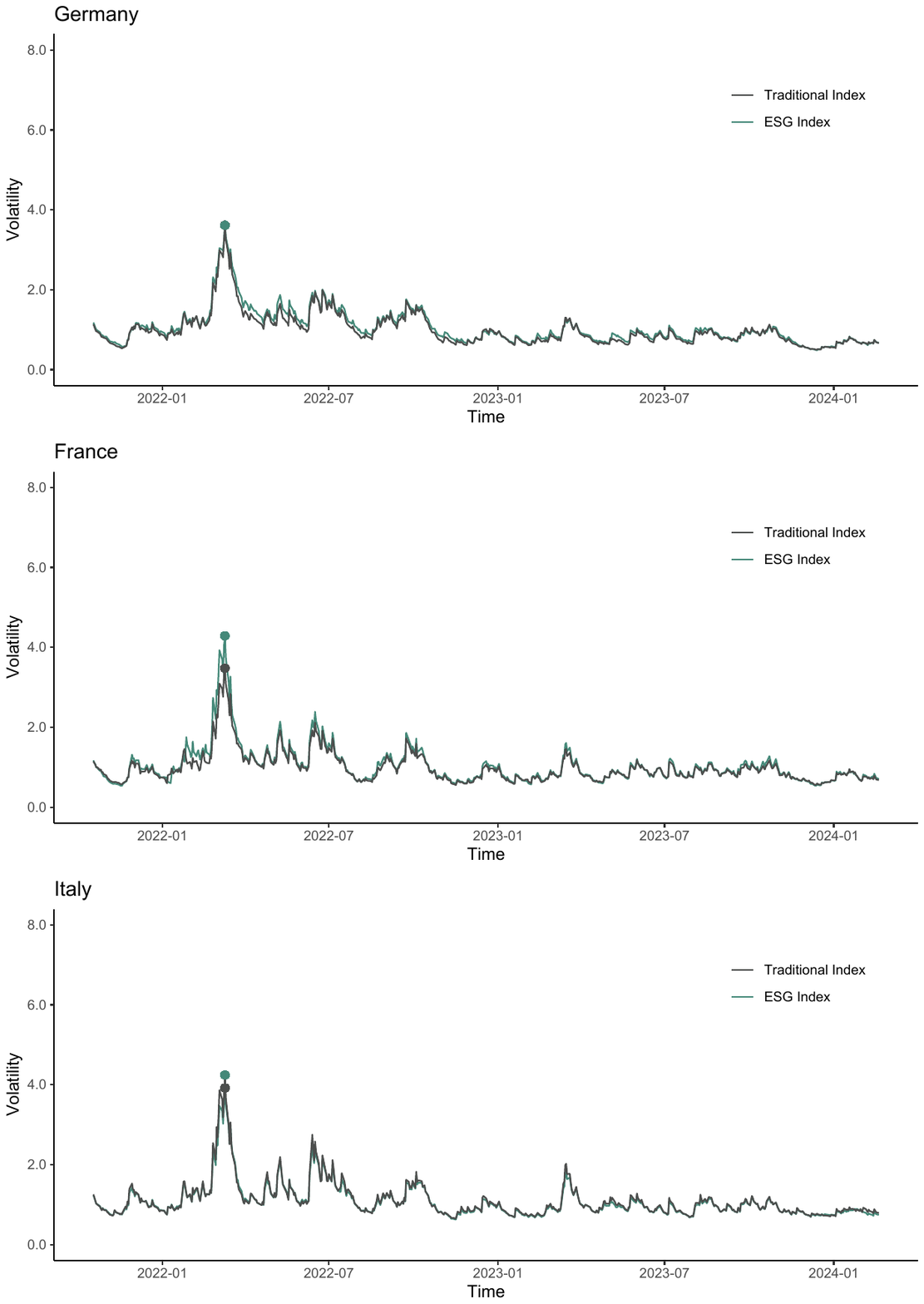}
\caption{Estimated volatility by index type and market.}\label{fig:vol_plot}	
\end{figure}

For a complete risk assessment, it is essential to examine the estimated conditional volatility. Figure \ref{fig:vol_plot} represents the estimated volatility by index type and market. In Germany, the volatility between the ESG and traditional indices is almost indistinguishable. During the financial crisis triggered by the Russia-Ukraine conflict, the ESG index exhibited a peak volatility of 3.62, nearly identical to the traditional index, which peaked at 3.61. France, on the other hand, presents a contrasting pattern. The ESG index showed notably higher volatility than the traditional index. During the same crisis period, the ESG index reached a peak volatility of 4.28, significantly exceeding the traditional index, which peaked at 3.47. Italy, like Germany, showed only a slight difference in volatility between the ESG and traditional indices. During the financial crisis, the ESG index recorded a peak volatility of 4.24, only marginally higher than the traditional index, which peaked at 3.91. ESG and traditional indices often behave similarly in terms of volatility, as seen in Germany and Italy, there are exceptions, such as in France, where ESG investments appear more volatile during periods of financial instability.

\section{Discussion and conclusions}\label{sec.conclusions}

Throughout the period affected by the COVID-19 emergency, the strong turbulence in the financial markets highlighted the resilience of ESG assets compared to the rest of the market. This demonstrated that ethics and financial performance can coexist, contradicting previous claims that they represented an added burden. However, it is important to emphasise that this is a complex relationship and that ethical investments do not always produce the expected financial returns \citep{burghof2021investing}. Ethical investments should not be mistaken for completely risk-free, as they remain susceptible to negative market shocks \citep{diaz2016return}.

Over the years, various scholars have shown that responsible investing is not synonymous with philanthropy. Instead, it reflects a shift in investors' priorities, where they focus not only on profitability but also on sustainability and responsible behaviour. Increasingly, investors are choosing to direct their capital towards companies that are committed to making the world a better place to live \citep{Strampelli_2022}. Ethical and responsible investing is important to align personal wealth management with broader social values, promoting a more sustainable and responsible investment approach. There is a need for a deeper understanding of investor psychology, particularly in the context of a global emergency such as COVID-19 \cite{hii2023behavioural}.

Conducting studies on these topics can provide valuable insight for companies and enable them to respond appropriately to future challenges. During periods of deep crisis, such as the recent Covid-19 pandemic or the last war conflicts, it is necessary to understand what the investors' behaviour is, if their choices tend to shift toward sustainable investment strategies or not. Consequently, investors, policymakers, and scholars should work together to identify and promote sustainable investment practices and strategies that consider ESG factors.
Thus, this research confirms that periods of turbulence affected volatility and financial returns, as revealed in previous studies \citep{Duttilo2023,mythesis}. Specifically, the first part of this study confirms that the performance of the ESG indices and that of traditional indices register both volatility in times of crisis \textbf{(H1)}. The findings of this study indicate that not only the pandemic phenomenon but also the war conflicts, particularly that in Ukraine started in 2022, had a significant impact on investments in Europe, amplifying the volatility of financial instruments. European financial markets were influenced by both the demand and supply sides of general investments. Ongoing conflicts, including those in the Middle East, have challenged the sustainability of certain sectors, particularly energy, where oil and gas companies saw a reevaluation of their role. This was a direct consequence of the increase in the demand for traditional energy sources to address the energy crisis. This situation created tension between the immediate need for energy security and the long-term goals of transitioning to renewable energy, leading to further fluctuations in financial markets. Social inequalities were exacerbated, resulting in humanitarian crises, leading investors to re-assess the risks associated with investments in regions impacted by extreme events or companies indirectly involved, leading some investors to divest from businesses and countries perceived as too risky, thus increasing the volatility of financial portfolios. Finally, governance was also affected by wars, as companies were forced to take positions on their operations in conflict regions and reconsider policies on issues such as human rights and sustainability, which had repercussions on stock prices.

However, the goal of this investigation is to discover if ESG portfolios are characterised by low risks compared to conventional portfolios; therefore, ESG indices show minor volatility compared to traditional indices \textbf{(H2)}. In previous parts, it emerged that ESG investments represent a safe haven during significant upheaval and crisis periods. They offer investors a means of managing risk and achieving positive returns while promoting economic sustainability
\citep{Katsampoxakis2024,Lin2024,risks11100182}.

Western European countries like Germany have shown a stronger inclination toward ESG investments. In these regions, the risk-return profile of ESG investments tends to be more favourable compared to traditional investments, primarily due to economic stability and government policies that encourage sustainability. Volatility has been relatively low compared to traditional investments, as many companies in these regions have integrated ESG practices into their business strategies for years, reducing the risks associated with environmental and social factors. In addition, benefiting from well-developed financial markets, they offer greater liquidity and better access to high-quality information, helping stabilise returns. Italy also showed results aligned with Germany. After an initial spike in the ESG indices at the outbreak of the Russia-Ukraine war, the situation then returned to calm, showing stability and alignment in the turbulence with the traditional indices. Therefore, the two different indices, ESG and traditional, tend to have the same level of volatility. On the other hand, France, probably unexpectedly, shows a spike in volatility on ESG stocks, at the outbreak of war, much more pronounced than in Italy. In subsequent periods, the situation seems to return almost to alignment with the other securities, but ESG securities, even if only slightly, show more volatility than traditional ones. Ultimately, H1 can be confirmed, that the volatility recorded, at least in the three countries investigated, Germany, France and Italy, tends to be  the same during a moment of crisis, such as that of the outbreak of a war. 
 
Therefore, H2, which wanted to confirm a lower volatility of the ESG indices in relation to traditional indices, was not confirmed. France notes a peak of greater volatility in the ESG indices, even if it is attenuating over time but remains greater for traditional indices. Italy and Germany, albeit at different times, also report higher volatility on ESG securities, but this is subsequently realigned.

The reasons for the higher volatility of ESG indices may refer to the fact that ESG practices are more exposed to economic, political, and social fluctuations. The post-pandemic economic crisis and the dependence on more vulnerable sectors, such as tourism and agriculture, have increased the risks associated with investments. This is why the risk-return profile of ESG investments tends to be less favorable compared to traditional investments, especially in sectors less resilient to global crises.
In this way, the reaction of stock returns is dynamic, and it shows that in each country, traditional indices consistently report higher mean returns than ESG indices, with Italy showing the largest gap \textbf{(H3)}. But, this deviation, is slightly
higher for ESG indices in each country, thus the differences are minimal.
In conclusion, it is possible to argue these last results: Germany has faced a series of economic, social, and political challenges that have redefined its approach to sustainability and ESG investments. The energy crisis, inflation, internal political tensions, and geopolitical dynamics have deeply influenced the investment landscape, creating opportunities and significant risks. Despite these difficulties, Germany has continued to be a leader in the green transition, with a strong commitment to sustainable innovation and a pragmatic approach to crisis management, making it a crucial market for ESG investors in Europe.

France has also faced a period of economic, social, and political challenges that have significantly impacted the investment landscape, including ESG investments. Social unrest, the energy crisis, stricter sustainability policies and political uncertainties have contributed to a highly volatile environment that requires a more cautious and well-informed approach for investors interested in the French market.

Finally, Italy has also faced a period of great challenges and transformations, which have had a significant impact on the investment context. The complexity of the period has created an environment characterised by high volatility and uncertainty, but also by opportunities for investments, particularly in sustainability and innovation. Despite the difficulties, Italy has remained a key player in Europe, striving to balance the needs for economic growth with its commitment to a green and sustainable transition. For example, structural reforms related to the National Recovery and Resilience Plan (PNRR) have had a significant impact on investor confidence and the perception of risk associated with investments in Italy. PNRR has provided crucial support to the economy, helping to strengthen ESG investments in sectors such as green infrastructure and digitalisation. These are the reasons why the situation that emerged in Italy is more interesting in this field and it reveals aligned reactions: ESG indices and traditional indices.

In the previous part, it was also analysed that one possible way, as a suggested solution, to operate in financial turmoil is to reduce uncertainty through the improvement of intangibles: for instance, a good reputation building. A strong reputation may help companies recover from market volatility better than a weaker reputation. It was confirmed during the October 1997 market turmoil, resulting in a significant role of reputation in stock performance \citep{Gregory1998}. Although the results of this research opened only a small vision in this field, it may represent a point of attention for future research, based on the power of intangibles in financial turmoil situations. Thus, more research should verify these results considering also the role of different kinds of crises and to understand better the role of reputation in avoiding the collapse of titles and consequently, of companies.

The findings of this study contribute to the growing body of research on the positive relationships between ethics and performance while also encouraging further studies on similar topics, using statistical investigations with indices different from those used in this study, to verify the trends that have emerged.

The limitations of this research are mainly related to the number of countries studied. France, Germany, and Italy. Although these are the top three EU countries by GDP, at least during the 2021-2024 period, the effects may differ from the results in other countries.

\bibliographystyle{apalike}
\bibliography{References.bib}

\begin{thebibliography}{}

\bibitem[Abate et~al., 2023]{Abate2023}
Abate, G., Basile, I., and Ferrari, P. (2023).
\newblock The integration of environmental, social and governance criteria in
  portfolio optimization: An empirical analysis.
\newblock {\em Corporate Social Responsibility and Environmental Management},
  31.

\bibitem[Ahlstr\"{o}m and Monciardini, 2022]{Ahlstrom2022}
Ahlstr\"{o}m, H. and Monciardini, D. (2022).
\newblock The regulatory dynamics of sustainable finance: Paradoxical success
  and limitations of eu reforms.
\newblock {\em Journal of Business Ethics}, 177.

\bibitem[Ahmad et~al., 2024]{ahmad2024environmental}
Ahmad, H., Yaqub, M., and Lee, S.~H. (2024).
\newblock Environmental, social, and governance related factors for business
  investment and sustainability: A scientometric review of global trends.
\newblock {\em Environment, Development and Sustainability}, 26(2):2965--2987.

\bibitem[Bauer and Smeets, 2015]{Bauer2015}
Bauer, R. and Smeets, P. (2015).
\newblock Social identification and investment decisions.
\newblock {\em Journal of Economic Behavior \& Organization}, 117(C):121--134.

\bibitem[Bauer et~al., 2007]{Bauer2007}
Bauer, T., Bodner, T., Erdogan, B., Truxillo, D., and Tucker, J. (2007).
\newblock Newcomer adjustment during organizational socialization: A
  meta-analytic review of antecedents, outcomes, and methods.
\newblock {\em The Journal of applied psychology}, 92:707--21.

\bibitem[Benlemlih and Bitar, 2018]{Benlemlih2018}
Benlemlih, M. and Bitar, M. (2018).
\newblock Corporate social responsibility and investment efficiency.
\newblock {\em Journal of Business Ethics}, Forthcoming.

\bibitem[Blajer-Golebiewska, 2014]{blajer2014corporate}
Blajer-Golebiewska, A. (2014).
\newblock Corporate reputation and economic performance: The evidence from
  poland.
\newblock {\em Economics \& Sociology}, 7(3):194.

\bibitem[Brooks and Oikonomou, 2018]{BROOKS20181}
Brooks, C. and Oikonomou, I. (2018).
\newblock The effects of environmental, social and governance disclosures and
  performance on firm value: A review of the literature in accounting and
  finance.
\newblock {\em The British Accounting Review}, 50(1):1--15.
\newblock The Effects of Environmental, Social and Governance Disclosures and
  Performance on Firm Value.

\bibitem[Burghof and Gehrung, 2021]{burghof2021investing}
Burghof, H.~P. and Gehrung, M. (2021).
\newblock Investing ethical: Harder than you think.
\newblock In {\em Ethical Discourse in Finance: Interdisciplinary and Diverse
  Perspectives}, pages 149--168. Springer.

\bibitem[Capelli et~al., 2021]{capelli2021forecasting}
Capelli, P., Ielasi, F., and Russo, A. (2021).
\newblock Forecasting volatility by integrating financial risk with
  environmental, social, and governance risk.
\newblock {\em Corporate Social Responsibility and Environmental Management},
  28(5):1483--1495.

\bibitem[Chatzitheodorou et~al., 2019]{Chatzitheodorou2019}
Chatzitheodorou, K., Skouloudis, A., Evangelinos, K., and Nikolaou, I. (2019).
\newblock Exploring socially responsible investment perspectives: A literature
  mapping and an investor classification.
\newblock {\em Sustainable Production and Consumption}, 19.

\bibitem[Chen et~al., 2023]{chen2023environmental}
Chen, S., Song, Y., and Gao, P. (2023).
\newblock Environmental, social, and governance (esg) performance and financial
  outcomes: Analyzing the impact of esg on financial performance.
\newblock {\em Journal of Environmental Management}, 345:118829.

\bibitem[Cole et~al., 2014]{cole2014applying}
Cole, S., Brown, M., and Sturgess, B. (2014).
\newblock Applying reputation data to enhance investment performance.
\newblock {\em World Economics}, 15(4):59--72.

\bibitem[Cowton, 1999]{Cowton1999}
Cowton, C. (1999).
\newblock Playing by the rules: ethical criteria at an ethical investment fund.
\newblock {\em Business Ethics: A European Review}, 8(1):60--69.

\bibitem[Demers et~al., 2021a]{demers2021esg}
Demers, E., Hendrikse, J., Joos, P., and Lev, B. (2021a).
\newblock Esg did not immunize stocks during the covid‐19 crisis, but
  investments in intangible assets did.
\newblock {\em Journal of Business Finance \& Accounting}, 48(3-4):433--462.

\bibitem[Demers et~al., 2021b]{Demers2021}
Demers, E., Hendrikse, J., Joos, P., and Lev, B. (2021b).
\newblock Esg didn't immunize stocks during the covid‐19 crisis, but
  investments in intangible assets did.
\newblock {\em Journal of Business Finance \& Accounting}, 48.

\bibitem[Diaz, 2016]{diaz2016return}
Diaz, J. F.~T. (2016).
\newblock Return and volatility performance comparison of ethical and
  non-ethical publicly-listed financial services companies.
\newblock {\em Ética, economía y bienes comunes}, 13(1).

\bibitem[Duttilo, 2024]{mythesis}
Duttilo, P. (2024).
\newblock {\em Modelling finacial returns with mixture of generalized normal
  distributions}.
\newblock Phd thesis, University "G. d'Annunzio" of Chieti-Pescara, Pescara,
  IT.
\newblock Available at \url{https://doi.org/10.48550/arXiv.2411.11847}.

\bibitem[Duttilo et~al., 2021]{Duttilo2021}
Duttilo, P., Gattone, S.~A., and Di~Battista, T. (2021).
\newblock Volatility modeling: An overview of equity markets in the euro area
  during covid-19 pandemic.
\newblock {\em Mathematics}, 9(11).

\bibitem[Duttilo et~al., 2023]{Duttilo2023}
Duttilo, P., Gattone, S.~A., and Iannone, B. (2023).
\newblock Mixtures of generalized normal distributions and egarch models to
  analyse returns and volatility of esg and traditional investments.
\newblock {\em AStA Adv Stat Anal}, pages 1--33.

\bibitem[Fich et~al., 2015]{FICH201521}
Fich, E.~M., Harford, J., and Tran, A.~L. (2015).
\newblock Motivated monitors: The importance of institutional investors'
  portfolio weights.
\newblock {\em Journal of Financial Economics}, 118(1):21--48.

\bibitem[Fombrun, 2012]{Fombrun2012}
Fombrun, C. (2012).
\newblock The building blocks of corporate reputation: Definitions,
  antecedents, consequences.
\newblock {\em The Oxford Handbook of Corporate Reputation}, pages 94--113.

\bibitem[Fombrun et~al., 2013]{Fombrun2013}
Fombrun, C., Gardberg, N., and Sever, J. (2013).
\newblock The reputation quotientsm: A multi-stakeholder measure of corporate
  reputation.
\newblock {\em Journal of Brand Management}, 7.

\bibitem[Friede, 2019]{Friede2019}
Friede, G. (2019).
\newblock Why don't we see more action? a metasynthesis of the investor
  impediments to integrate environmental, social, and governance factors.
\newblock {\em Business Strategy and the Environment}, 28(6):1260--1282.

\bibitem[Fung et~al., 2024]{fung2024impact}
Fung, J.~K., Lam, F.~E., and Tse, Y. (2024).
\newblock The impact of esg rating on hedging downside risks: Evidence from a
  weight-tilted hang seng index.
\newblock {\em Journal of Risk and Financial Management}, 17(2):57.

\bibitem[Garrido~Ruso et~al., 2024]{Garrido2024}
Garrido~Ruso, M., González, L., López-Penabad, M.-C., and Santomil, P.
  (2024).
\newblock Does esg implementation influence performance and riskin smes?
\newblock {\em Corporate Social Responsibility and Environmental Management},
  31.

\bibitem[Ghani et~al., 2024]{ghani2024forecasting}
Ghani, U., Zhu, B., Qin, Q., and Ghani, M. (2024).
\newblock Forecasting us stock market volatility: Evidence from esg and cpu
  indices.
\newblock {\em Finance Research Letters}, 59:104811.

\bibitem[Godfrey et~al., 2009]{godfrey2009relationship}
Godfrey, P.~C., Merrill, C.~B., and Hansen, J.~M. (2009).
\newblock The relationship between corporate social responsibility and
  shareholder value: An empirical test of the risk management hypothesis.
\newblock {\em Strategic Management Journal}, 30:425--445.

\bibitem[Gregory, 1998]{Gregory1998}
Gregory, J.~R. (1998).
\newblock Does corporate reputation provide a cushion to companies facing
  market volatility? some supportive evidence.
\newblock {\em Corporate Reputation Review}, 1(3):288--290.

\bibitem[Gupta and Chaudhary, 2023]{risks11100182}
Gupta, H. and Chaudhary, R. (2023).
\newblock An analysis of volatility and risk-adjusted returns of esg indices in
  developed and emerging economies.
\newblock {\em Risks}, 11(10).

\bibitem[Han et~al., 2022]{HAN2022}
Han, Y., Huang, J., Li, R., Shao, Q., Han, D., Luo, X., and Qiu, J. (2022).
\newblock Impact analysis of environmental and social factors on early-stage
  covid-19 transmission in china by machine learning.
\newblock {\em Environmental Research}, 208:112761.

\bibitem[Hemerijck, 2018]{Hemerijck2018}
Hemerijck, A. (2018).
\newblock Social investment as a policy paradigm.
\newblock {\em Journal of European Public Policy}, 25(6):810--827.

\bibitem[Hii et~al., 2023]{hii2023behavioural}
Hii, I.~S., Li, X., and Zhu, H. (2023).
\newblock Behavioural biases and investment decisions during covid-19: An
  empirical study of chinese investors.
\newblock {\em Institutions and Economies}, pages 81--103.

\bibitem[Katsampoxakis et~al., 2024]{Katsampoxakis2024}
Katsampoxakis, I., Xanthopoulos, S., Basdekis, C., and Christopoulos, A.~G.
  (2024).
\newblock Can esg stocks be a safe haven during global crises? evidence from
  the covid-19 pandemic and the russia-ukraine war with time-frequency wavelet
  analysis.
\newblock {\em Economies}, 12(4).

\bibitem[Krosinsky and Robins, 2008]{krosinsky2008}
Krosinsky, C. and Robins, N. (2008).
\newblock {\em Sustainable Investing: The Art of Long-term Performance}.
\newblock Environmental Market Insights Series. Earthscan.

\bibitem[Lapanan, 2018]{Lapanan2018}
Lapanan, N. (2018).
\newblock The investment behavior of socially responsible individual investors.
\newblock {\em The Quarterly Review of Economics and Finance}, 70(C):214--226.

\bibitem[Lin and Bali~Swain, 2024]{Lin2024}
Lin, X. and Bali~Swain, R. (2024).
\newblock Performance of negatively screened sustainable investments during
  crisis.
\newblock {\em International Review of Economics \& Finance}, 93.

\bibitem[Michelucci, 2016]{michelucci2016}
Michelucci, F.~V. (2016).
\newblock Social impact investments: Does an alternative to the anglo-saxon
  paradigm exist?
\newblock {\em Voluntas}, 28(6):1--24.

\bibitem[Miras et~al., 2020]{Miras2020}
Miras, M. d.~M., Bravo, F., and Bernabé, E. (2020).
\newblock Does corporate social responsibility reporting actually destroy firm
  reputation?
\newblock {\em Corporate Social Responsibility and Environmental Management},
  27.

\bibitem[Mohamed and Pijourlet, 2017]{Mohamed2017}
Mohamed, A. and Pijourlet, G. (2017).
\newblock Csr performance and the value of cash holdings: International
  evidence.
\newblock {\em Journal of Business Ethics}, DOI: 10.1007/s10551-015-2658-5.

\bibitem[Nawrocki and Szwajca, 2022]{nawrocki2022importance}
Nawrocki, T.~L. and Szwajca, D. (2022).
\newblock The importance of selected aspects of a company’s reputation for
  individual stock market investors—evidence from polish capital market.
\newblock {\em Sustainability}, 14(15):9187.

\bibitem[Negara et~al., 2024]{negara2024impact}
Negara, N. G.~P., Ishak, G., and Priambodo, R. E.~A. (2024).
\newblock The impact of esg disclosure score on firm value: Empirical evidence
  from esg listed company in indonesia stock exchange.
\newblock {\em European Journal of Business and Management Research},
  9(2):114--118.

\bibitem[Oehmke and Opp, 2021]{Oehmke2021}
Oehmke, M. and Opp, M. (2021).
\newblock {A theory of socially responsible investment}.
\newblock LSE Research Online Documents on Economics 118891, London School of
  Economics and Political Science, LSE Library.

\bibitem[Price and Sun, 2017]{price2017doing}
Price, J.~M. and Sun, W. (2017).
\newblock Doing good and doing bad: The impact of corporate social
  responsibility and irresponsibility on firm performance.
\newblock {\em Journal of Business Research}, 80:82--97.

\bibitem[Rehman et~al., 2022]{Rehman2022}
Rehman, S.~U., Bresciani, S., Yahiaoui, D., and Giacosa, E. (2022).
\newblock Environmental sustainability orientation and corporate social
  responsibility influence on environmental performance of small and medium
  enterprises: The mediating effect of green capability.
\newblock {\em Corporate Social Responsibility and Environmental Management},
  29(6):1954--1967.

\bibitem[Rizzi et~al., 2018]{rizzi2018}
Rizzi, F., Pellegrini, C., and Battaglia, M. (2018).
\newblock The structuring of social finance: Emerging approaches for supporting
  environmentally and socially impactful projects.
\newblock {\em Journal of Cleaner Production}, 170:805--817.

\bibitem[Salzmann, 2013]{salzmann2013}
Salzmann, A.~J. (2013).
\newblock The integration of sustainability into the theory and practice of
  finance: An overview of the state of the art and outline of future
  developments.
\newblock {\em Journal of Business Economics}, 83(6):555--576.

\bibitem[Sch\"{u}rmann, 2006]{schurmann2006reputation}
Sch\"{u}rmann, S. (2006).
\newblock Reputation: Some thoughts from an investor's point of view.
\newblock {\em The Geneva Papers on Risk and Insurance-Issues and Practice},
  31:454--469.

\bibitem[Sharma et~al., 2024]{Sharma2024}
Sharma, S., Aggarwal, V., and Mehta, G. (2024).
\newblock Esg performance and corporate volatility: an empirical exploration in
  an emerging economy.
\newblock {\em International Journal of Social Economics}.

\bibitem[Strampelli, 2022]{Strampelli_2022}
Strampelli, G. (2022).
\newblock {\em Institutional Investor Stewardship in Italian Corporate
  Governance}, page 130–149.
\newblock Cambridge University Press.

\bibitem[Tekula and Shah, 2016]{tekula2016}
Tekula, R. and Shah, A. (2016).
\newblock Impact investing: Funding social innovation.
\newblock In Lehner, O.~M., editor, {\em Routledge Handbook of Social and
  Sustainable Finance}. Routledge.

\bibitem[{University of Cambridge}, 2022]{UniCam2022}
{University of Cambridge} (2022).
\newblock What is responsible investment?
\newblock Accessed: March 16, 2022.

\bibitem[Weber, 2016]{weber2016}
Weber, O. (2016).
\newblock Introducing impact investing.
\newblock In Lehner, O.~M., editor, {\em Routledge Handbook of Social and
  Sustainable Finance}. Routledge, Taylor Francis Group.

\bibitem[Wen et~al., 2022]{Wen2020}
Wen, L., Qiu, Y., Wang, M., Yin, J., and Chen, P. (2022).
\newblock Numerical characteristics and parameter estimation of finite mixed
  generalized normal distribution.
\newblock {\em Communications in Statistics - Simulation and Computation},
  51(7):3596--3620.

\bibitem[Zanatto et~al., 2023]{Zanatto2023}
Zanatto, C., Catalão-Lopes, M., Pina, J.~P., and Carrilho-Nunes, I. (2023).
\newblock The impact of esg news on the volatility of the portuguese stock
  market—does it change during recessions?
\newblock {\em Business Strategy and the Environment}, 32(8):5821--5832.

\bibitem[Zappa, 1937]{zappa1937reddito}
Zappa, G. (1937).
\newblock {\em Il reddito di impresa. Scritture doppie, conti e bilanci di
  aziende commerciali}.
\newblock Giuffrè, 2 edition.

\bibitem[Zappa, 1958]{zappa1958}
Zappa, G. (1958).
\newblock L'ipotesi del costante nella dottrina e nella gestione di azienda.
\newblock {\em Il Risparmio}, 6(12):2185--2199.

\bibitem[Zappa, 1959]{zappa1959}
Zappa, G. (1959).
\newblock La perdurante instabilità dei mercati e delle gestioni di azienda.
\newblock {\em Il Risparmio}, 7(2):163--184.

\bibitem[Zappa, 1960]{zappa1960}
Zappa, G. (1960).
\newblock Il divenire sociale.
\newblock {\em Il Risparmio}, 8(1):1--35.

\bibitem[Zigrand, 2014]{Zigrand2014}
Zigrand, J.-P. (2014).
\newblock {Systems and systemic risk in finance and economics}.
\newblock LSE Research Online Documents on Economics 61220, London School of
  Economics and Political Science, LSE Library.

\end{thebibliography}

\textbf{Disclosure statement} The authors report that there are no competing interests to declare.

\appendix
\setcounter{table}{0}
\renewcommand{\thetable}{A\arabic{table}}

\section{Tables}\label{app:tables}

\begin{table}[ht]
\centering
\begin{minipage}{\textwidth}
\caption{Estimated basic statistics of traditional and ESG indices}
\label{tab:BasicStatistics}
\begin{tabular}{lcccccccc}
\toprule
& \multicolumn{2}{c}{Mean}&  \multicolumn{2}{c}{Stdev}&  \multicolumn{2}{c}{Skew}&  \multicolumn{2}{c}{Kur}\\
&\multicolumn{1}{c}{Trad.} & \multicolumn{1}{c}{ESG} &
\multicolumn{1}{c}{Trad.} & \multicolumn{1}{c}{ESG} &
\multicolumn{1}{c}{Trad.} & \multicolumn{1}{c}{ESG} &
\multicolumn{1}{c}{Trad.} & \multicolumn{1}{c}{ESG}\\
\midrule
Germany & 0.0166  & 0.0031 & 1.1487   &   1.1989 &0.0840  &   0.1670& 4.4968    &  4.4452\\
France &0.0254 &  0.0229 & 1.1396  & 1.1721&  -0.0896 & -0.0069 &3.9892 &   4.8177\\
Italy & 0.0313 &  0.0289&  1.2633 &  1.2972& -0.5537  &-0.6237& 3.7299   &4.0547\\
\bottomrule  
\end{tabular}
\end{minipage}
\end{table}

\begin{table}[ht]
\caption{Statistical tests results}\label{tab:StatisticalTestsResults}
\centering 
\resizebox{9cm}{!}{
\begin{tabular}{lrrr}
\toprule
Index & JB test & ADF test  & ARCH-LM test\\
\midrule
DAX & 510.20** &  -17.73** &  61.40**\\
DAX30ESGK & 500.68** & -17.20** &  57.70**\\
CAC40 & 401.99** & -17.93** &  54.70**\\
CAC40ESG & 584.64** & -17.56** &  52.00**\\
FTSEMIB & 381.57** & -17.70** &  47.50**\\
MIBESG & 453.40** & -17.61** &  48.60**\\
\bottomrule 
\end{tabular}
}
\begin{tablenotes}
\centering
\item[]{\footnotesize Note. ** denotes the significance of the p-value at the level 1\%.}
\end{tablenotes}
\end{table}

\begin{table}[ht]
\caption{Estimated parameters of the two-component MGND model}
\label{tab:MixtureofGND}
\centering 
\resizebox{12cm}{!}{
\begin{tabular}{lccrrcccccccccc}
\toprule
\multicolumn{4}{c}{Stable component}&  \multicolumn{4}{c}{Turmoil component}\\
\multicolumn{1}{c}{$\pi_1$} & \multicolumn{1}{c}{$\mu_1$} & \multicolumn{1}{c}{$\delta_1$} & \multicolumn{1}{c}{$\nu_1$}&  \multicolumn{1}{c}{$\pi_2$} & \multicolumn{1}{c}{$\mu_2$} & \multicolumn{1}{c}{$\delta_2$} & \multicolumn{1}{c}{$\nu_2$}\\
\midrule
0.8502 & 0.1141 & 0.7351 & 1.3123 &  0.1498 & -0.4408 & 2.0158 & 1.7998\\
\bottomrule 
\end{tabular}
}
\end{table}

\begin{table}[ht]
\caption{Estimated coefficients of the conditional mean equation}
\label{tab:Conditional Mean}
\centering
\resizebox{10cm}{!}{
\begin{tabular}{llll}
\toprule
Index & \multicolumn{1}{c}{$\mu$} & \multicolumn{1}{c}{$m_1$}  & \multicolumn{1}{c}{$\lambda_1$}\\
\midrule
DAX & -0.0887* & -2.6587** & 0.2083** \\
DAX30ESGK & -0.0404 & -2.6888** & 0.1428**\\
CAC40 & -0.1690 & -2.6935** & 0.3117** \\
CAC40ESG & -0.0633** & -2.2656** & 0.1713** \\
FTSEMIB & -0.1476** & -2.7659** & 0.2795**\\
MIBESG & -0.1461** & -2.7715** & 0.2704**\\
\bottomrule
\end{tabular}
}
\begin{tablenotes}
\centering
\item[]{\footnotesize Note. ** and * denote the significance of the p-value at the level of 1\% and 5\% level, respectively.}
\end{tablenotes}
\end{table}

\begin{table}[ht]
\caption{Estimated coefficients of the conditional volatility equation}
\label{tab:Conditional Variance}
\centering 
\resizebox{16cm}{!}{
\begin{tabular}{llllllll}
\toprule
Index & \multicolumn{1}{c}{$\omega$} & \multicolumn{1}{c}{$v_1$} & \multicolumn{1}{c}{$\alpha_1$} & \multicolumn{1}{c}{$\gamma_1$}& \multicolumn{1}{c}{$\beta_1$}& \multicolumn{1}{c}{$\nu$} & \multicolumn{1}{c}{$s$}\\
\midrule
DAX & -0.0320* &  0.4841** & -0.1123** & 0.1121* & 0.9389** & 1.4890** &1.0295** \\
DAX30ESGK & -0.0215* & 0.3940** & -0.1051** & 0.1147** & 0.9536** &  1.3553** & 1.0330** \\
CAC40 & -0.0435** & 0.5934** & -0.1420** & 0.0546 & 0.9105** & 1.6868** & 1.0558**\\
CAC40ESG & -0.0384** & 0.6434** & -0.1554** & 0.0798** & 0.9102** & 1.7020** & 1.0066**\\
FTSEMIB & -0.0255* & 0.6166** & -0.1129** & 0.0863** & 0.9026** & 1.8300** & 0.9999**\\
MIBESG & -0.0238* & 0.7009** & -0.1221** & 0.1024** & 0.8856** & 1.8833** & 0.9965**\\
\bottomrule
\end{tabular}
}
\begin{tablenotes}
\centering
\item[]{\footnotesize Note. ** and * denote the significance of the p-value at the level of 1\% and 5\% level, respectively.}
\end{tablenotes}
\end{table}


\end{document}